\documentclass[twocolumn,floatfix,showpacs,preprintnumbers,superscriptaddress,amsmath,amssymb]{revtex4}

\usepackage{graphicx}

\begin{document}

\title{Conductance of a hydrogen molecule}

\author {J. Heurich} 
\affiliation{Institut f\"ur Theoretische Festk\"orperphysik, Universit\"at Karlsruhe, 
76128 Karlsruhe, Germany}

\author {F. Pauly} 
\affiliation{Institut f\"ur Theoretische Festk\"orperphysik, Universit\"at Karlsruhe, 
76128 Karlsruhe, Germany}

\author {J.C. Cuevas}
\affiliation{Institut f\"ur Theoretische Festk\"orperphysik, Universit\"at Karlsruhe, 
76128 Karlsruhe, Germany}

\author {W. Wenzel} 
\affiliation{Forschungszentrum Karlsruhe, Institut f\"ur Nanotechnologie,
76021 Karlsruhe, Germany}

\author {Gerd Sch\"on}
\affiliation{Institut f\"ur Theoretische Festk\"orperphysik, Universit\"at Karlsruhe, 
76128 Karlsruhe, Germany}
\affiliation{Forschungszentrum Karlsruhe, Institut f\"ur Nanotechnologie,
76021 Karlsruhe, Germany}

\date{\today}

\begin{abstract}
A recent experiment reported that a single hydrogen molecule can form a bridge 
between Pt electrodes, which has a conductance close to one quantum unit, carried 
by a single channel. We present density functional calculations explaining these 
experimental findings. We show that the symmetry of the molecular orbitals selects
a single conduction channel. The transmission of this channel is close to unity due to a 
combination of charge transfer between H$_2$ and the Pt contacts and the strong 
hybridization between the bonding state of the molecule and the $d$ band of the Pt 
leads.
\end{abstract}


\maketitle

The recent advances in nanofabrication have triggered the hope that electronic 
devices can be shrunk down to the single molecule scale~\cite{Aviram1998,Joachim2000}.
Single molecules have been shown to perform functions analogous to those of the key 
microelectronics components such as switches~\cite{Gao2000,Collier1999,Reed2001},
rectifiers \cite{Metzger1999} and electronic mixers \cite{Chen1999}. However, the 
future of \emph{molecular electronics} depends crucially on our understanding
of the transport mechanisms in single-molecule junctions. From the experimental side
an unambiguous contact to a single molecule is difficult to achieve \cite{Cui2001}, 
and in many cases the measurements are not reproducible. On the other hand, the 
discrepancies between theory and experiment and even between different theories are 
notorious \cite{Ventra2000}.

From this perspective the measurement of the conductance of an individual hydrogen 
molecule reported by Smit \emph{et al.}~\cite{Smit2002} provides a valuable
opportunity to analyze the emerging concepts on the electrical conduction in 
single-molecule devices in the perhaps simplest possible system. In Ref.~\cite{Smit2002}
it was shown that a single hydrogen molecule can form a stable molecular bridge 
between platinum contacts. In contrast to results for organic molecules
\cite{Reed1997,Kergueris1999,Reichert2001}, this bridge has a conductance close 
to one quantum unit, carried by a single channel. This result belies the 
conventional wisdom because the closed-shell configuration of H$_2$ results in 
a huge gap between its bonding and antibonding states, making it a perfect
candidate for an insulating molecule. Thus, the observations of Ref.~\cite{Smit2002} 
pose an interesting theoretical challenge and their understanding can help to 
elucidate some of the basic transport mechanisms at the single molecule scale.

In this paper, we develop (i) a simple theoretical model to understand the conduction
mechanism through the hydrogen bridge which (ii) we match to quantitative density
functional theory (DFT) calculations of the conductance. The interplay of these
two approaches permits us to identify the two relevant mechanisms that give rise to
the large conductance of the hydrogen molecule. First, the catalytic activity of
platinum is responsible for a significant charge transfer between H$_2$ and the Pt 
contacts, which moves the bonding state of the molecule towards the Fermi energy.
Then, the strong hybridization with the $d$ band of the Pt leads provides a large
broadening of the bonding state, finally allowing for a high transmission. These
two key findings are at variance with the central conclusion of the theoretical
analysis presented in the experimental paper. Additionally, we show that, due to
the symmetry of its molecular orbitals, H$_2$ filters out only one of the Pt
conduction channels. We discuss the fundamental implications of these findings
for the proper description of the transport properties of single-molecule devices.

In order to investigate the transport through the hydrogen bridge reported in 
Ref.~\cite{Smit2002} we have performed DFT calculations of the linear conductance 
using the method described in Ref.~\cite{Heurich2002}, similar in spirit to 
Refs.~\cite{Yaliraki1999,Palacios2001}. In this approach, we first decompose the 
Hamiltonian of a molecular junction as $\hat{H} = \hat{H}_L + \hat{H}_R + 
\hat{H}_C + \hat{V}$, where $\hat{H}_C$ describes the ``central cluster" of the 
system, which contains the molecule plus part of the leads, $\hat{H}_{L,R}$ 
describe the left and right electrode respectively, and $\hat{V}$ gives the 
coupling between the electrodes and the central cluster. The electronic structure
of the central cluster is calculated within the DFT approximation~\cite{DFT}. The left 
and right reservoirs are modeled as two perfect bulk crystals of the corresponding 
metal using the tight-binding parameterization of Ref.~\cite{Mehl1996}. The
inclusion of part of the leads in the {\em ab initio} calculation assures the 
correct description of (i) the molecule-electrodes coupling, (ii) the charge 
transfer between the molecule and the leads and (iii) the lineup of the molecular 
levels relative to the metal Fermi level \cite{Xue2001}. The Fermi level is set
naturally by the highest occupied molecular orbital (HOMO) for a sufficiently 
large number of metallic atoms in the central cluster.

As in the experiment \cite{Smit2002} we concentrate on the analysis of the linear 
conductance, which is given by the Landauer formula $G = G_0 \mbox{Tr} \{ \hat{t} 
\hat{t}^{\dagger} \} = G_0 \sum_i T_i$, where $G_0=2e^2/h$ is the quantum of 
conductance, $\hat{t}$ is the transmission matrix and the $T_i$'s are the 
transmission eigenvalues at the Fermi energy $E_F$. The transmission matrix is given 
by $\hat{t}(\epsilon) = 2 \; \hat{\Gamma}^{1/2}_L(\epsilon) \hat{G}^r_{C}(\epsilon) 
\hat{\Gamma}^{1/2}_R(\epsilon)$ with the scattering rate matrices  
$\hat{\Gamma}_{L,R} = \mbox{Im} ( \hat{\Sigma}_{L,R} )$. Here $\hat{\Sigma}_{L,R}$ 
are the self-energies which contain the information of the electronic structure of 
the leads and their coupling to the central cluster. Finally, the Green functions of 
the central cluster are given by $\hat{G}_C(\epsilon) = \left[ \epsilon \hat{1} - 
\hat{H}_C - \hat{\Sigma}_L (\epsilon) - \hat{\Sigma}_R (\epsilon) \right]^{-1}$. 
For the case of non-orthogonal basis sets it is straightforward to implement the 
overlap matrix into these formulas via a L\"owdin transformation (see e.g. 
Ref.~\cite{Brandbyge1999}).

\begin{figure}[t]
\begin{center}
\includegraphics[width=0.85\columnwidth]{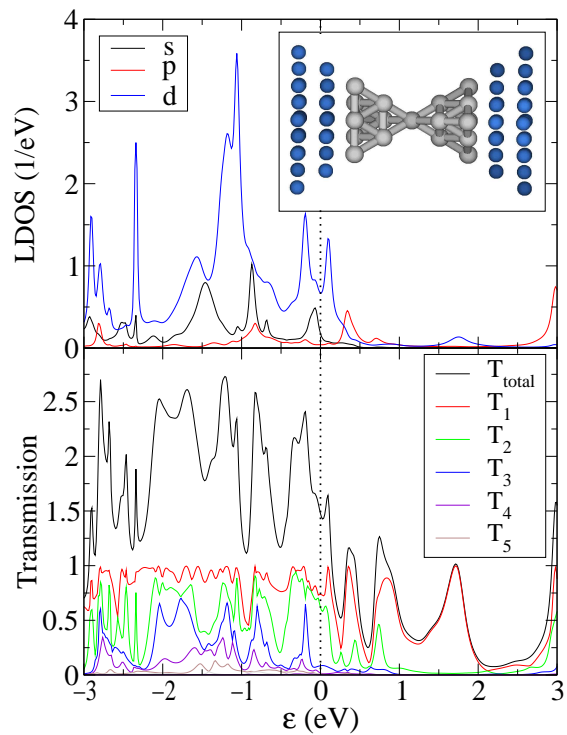}
\caption{\label{Pt} Transmission and LDOS at the central atom as a function of 
energy for the pure Pt structure shown in the inset. The LDOS is projected into the
Pt valence orbitals. The Fermi energy is set to zero. Inset: Pt one-atom contact 
consisting of a fcc structure with bulk interatomic distances grown along the (111) 
direction. The blue atoms represent the bulk atoms used to model the reservoirs.}
\end{center}
\end{figure}

In Ref.~\cite{Smit2002} a break junction was used to produce pure Pt contacts of 
atomic size. Then, it was shown that the transport through these contacts
was markedly altered in the presence of hydrogen. Following the experiment we first 
analyze the transport through a pure Pt single-atom contact. In the DFT calculations
we use the BP86 functional \cite{BP86} and the basis set of Christiansen \emph{et 
al.}~\cite{Ross1990}. For the description of the Pt reservoirs we use a basis with
the $5d,\,6s,\,6p$ orbitals. Fig.~\ref{Pt} shows the transmission and local density 
of states (LDOS)~\cite{note1} projected into the central atom of the structure 
shown in the inset. We find that the HOMO of the central cluster is at -5.44 
eV, which is close to the Pt work function $\sim \;$-5.64 eV. The gap between the
HOMO and the lowest unoccupied molecular orbital is almost closed (0.12 eV).
The total transmission at the Fermi energy is $T_{tot} = 1.47$ in agreement with 
the conductance histogram of Ref.~\cite{Smit2002}, which shows a large peak at 
1.4--1.8$G_0$. This transmission is made up of five conduction channels, which is
due to the contribution of the Pt $d$ orbitals. This is again a manifestation of the
fact that in single-atom contacts the number of channels is controlled by the valence
orbitals of the central atom \cite{Cuevas1998,Scheer1998}. In particular, the dominant
channel is a combination of the $s$ and $d_{z^2}$ orbitals \cite{Nielsen2002}.

\begin{figure}[t]
\begin{center}
\includegraphics[width=0.85\columnwidth]{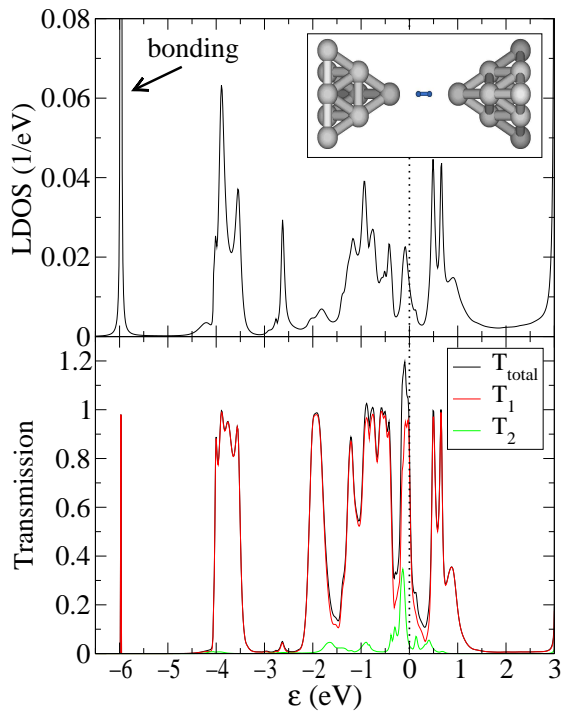}
\caption{\label{H2} Transmission and LDOS projected into one of the H atoms as a function
of energy for the Pt-H$_2$-Pt structure, the central cluster of which is shown in the 
inset. At $E_F$ $T_1=0.83$ and $T_2=0.03$. The H-H and Pt-H distances are 0.8 {\AA} 
and 2.1 {\AA} respectively. We use the cc-pVDZ basis set for H.}
\end{center}
\end{figure}

Let us now study how the presence of H$_2$ modifies the conduction. Usually the 
lack of knowledge of the precise geometry of the molecular junction complicates 
the comparison between theory and experiment. However, in Ref.~\cite{Smit2002} 
the presence of H$_2$ was identified by means of the signature of its vibrational 
modes in the conductance. This information establishes clear constraints on the 
geometries realized in the experiment. In this sense, we only consider 
configurations which are compatible with the observed vibrational modes.
The most probable configuration is shown in the inset of Fig.~\ref{H2}, where
the H$_2$ is coupled to a single Pt atom on either side (top position). In this
geometry the vibrational mode of the center of mass motion of H$_2$, which is
the one seen in the experiment, has an energy of 55.6 meV, lying in the range
of the experimental values. In Fig.~\ref{H2} we also show the transmission and 
the LDOS projected into the orbitals of one of the H atoms. The total transmission
at the Fermi energy is $T_{tot} = 0.86$ and it is largely dominated by a single
channel, in agreement with experimental results. We would like to draw the 
attention to the following two features in the LDOS: (i) the bonding state of 
the molecule appears as a peak at $\sim 6$ eV below $E_F$ and the antibonding state, 
not shown in Fig.~\ref{H2}, is located at $\sim 18$ eV above $E_F$. This 
indicates that the molecular character of H$_2$ is largely conserved. (ii) Around 
the Fermi energy the gap between the molecular states is filled due to the strong 
hybridization with the Pt leads.

\begin{figure}[t]
\begin{center}
\includegraphics[width=\columnwidth]{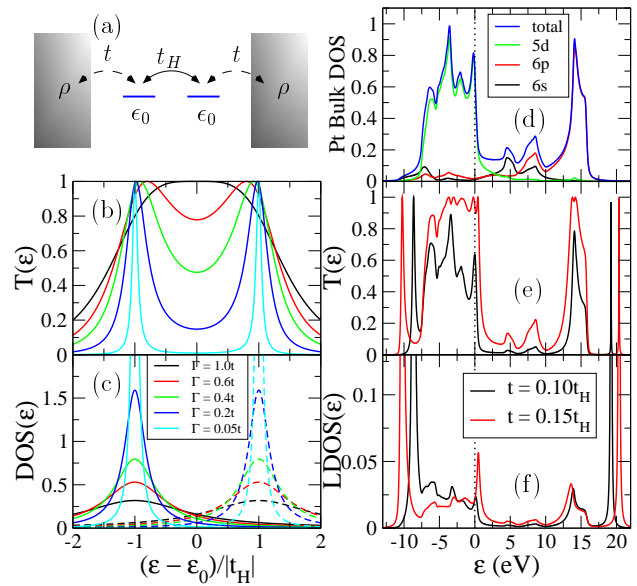}
\caption{\label{toymodel} (a) Schematic representation of the model described in the
text. (b-c) Transmission and DOS projected into the bonding (solid lines) and 
antibonding (dashed lines) states vs.~energy for different values of the scattering rate 
$\Gamma$. (d) Bulk DOS of the Pt reservoirs. (e-f) Transmission and LDOS projected into 
one of the H atoms for two values of the hopping $t$, assuming that the leads are modeled
by Pt bulk atoms, $t_H=$-12 eV and $\epsilon_0 = 6$ eV. In panels (d-f) the Fermi energy 
is set to zero.}
\end{center}
\end{figure}

We now proceed to explain the underlying physics with the help of a simple model. To 
illustrate the ingredients necessary to reproduce the full DFT results we shall progressively 
sophisticate the model.
We describe H$_2$ with a two-sites tight-binding model, see Fig.~\ref{toymodel}(a). 
In this scheme $\epsilon_0$ represents the 1$s$ energy level of H and $t_H$ is the hopping 
connecting the H atoms. This hopping is simply related to the splitting between the bonding 
($\epsilon_+$) and the antibonding state ($\epsilon_-$) of the molecule, namely 
$\epsilon_{\pm} = \epsilon_0 \pm t_H$, and its value is $\sim -12$ eV. The molecule is 
coupled symmetrically to the leads with a single hopping $t$. Obviously, within this model
the conductance is made up of a single channel, but let us postpone the discussion of this point
for the moment. The transmission is given by

\begin{equation}
T(\epsilon) =  \frac{4 \Gamma^2 t^2_H}{\left[(\epsilon - \tilde{\epsilon}_+)^2 
+ \Gamma^2 \right]  \left[(\epsilon - \tilde{\epsilon}_-)^2 + \Gamma^2 \right]}.
\end{equation}

\noindent
Here, $\tilde{\epsilon}_{\pm} = \epsilon_0 \pm t_H - t^2 \mbox{Re}\{g^a\}$ are the
renormalized molecular levels, $g^a(\epsilon)$ being the advanced Green function 
which describes the local electronic structure of the leads. The scattering rate 
$\Gamma$, which determines the broadening of the molecular levels, is given by 
$\Gamma(\epsilon) = t^2 \mbox{Im}\{g^a\} = \pi t^2 \rho$, where $\rho(\epsilon)$ 
is the LDOS of the Pt contacts. Let us first assume that $\Gamma$ is energy 
independent and that the levels are not renormalized $(\tilde{\epsilon}_{\pm} 
= \epsilon_{\pm})$. In Fig.~\ref{toymodel}(b) we show the transmission as a function
of energy for different values of $\Gamma$ in units of $t_H$. We also 
show in Fig.~\ref{toymodel}(c) the corresponding DOS projected into the bonding and 
antibonding states of H$_2$, which are given by $\rho_{\pm} = \Gamma / \pi \{ 
(\epsilon - \tilde{\epsilon}_{\pm})^2  + \Gamma^2 \}$ respectively. Taking into 
account the huge value of $t_H$, one might naively expect the curve for 
$\Gamma = 0.05 t_H$ to represent the relevant situation. Assuming that H$_2$
remains neutral, the bonding state is occupied by two electrons and $E_F = 
\epsilon_0$. In this simple picture H$_2$ would be insulating, in clear disagreement
with the observations of Ref.~\cite{Smit2002}. Our DFT calculations indicate that
there are two main ingredients missing in this simple argument. First,
H forms a covalent bond with Pt by sharing electrons. The DFT calculations show
that every H atom donates $\approx 0.12$e to Pt. This implies that the Fermi level 
lies closer to the bonding state. With this charge transfer the transmission raises 
significantly, but it is not yet sufficient to reproduce the DFT results. 
Thus, a large broadening ($\Gamma$) of the bonding state is still needed.
As suggested by the DFT calculations, this is provided by the good 
Pt-H$_2$ coupling and the large DOS around the Fermi energy coming from the $d$ 
band of Pt. We test this idea assuming that $g^a$ is the bulk Green function 
of Pt. The Pt bulk DOS is shown in Fig.~\ref{toymodel}(d). We also show in 
Fig.~\ref{toymodel}(e-f) the transmission and the LDOS projected into one of the 
H atoms for two values of the coupling to the leads $t$. One can see that for 
realistic values of $t \approx 1$-$2$ eV, the transmission at $E_F$ can now indeed reach 
values close to one. Therefore, we conclude that \emph{the high conductance
of H$_2$ is due to the charge transfer between H$_2$ and the Pt leads and the
strong hybridization between the bonding state and the $d$ band of Pt}. This 
mechanism is not exclusive of Pt and it must also operate in other transition 
metals, as it was shown experimentally for Pd (see Ref.~\cite{Smit2002}).

Let us now address why only a single channel is observed. In view of
the model described above a simple explanation could be that all the current flows
through the 1$s$ orbital of the closest H atom to the Pt atom. However, the DFT 
calculations show that while the Pt-Pt coupling is negligible, this is not the
case for the Pt and the second H atom. In principle there are two paths for the 
current, i.e. two channels may occur. Due to the spherical symmetry
of the H orbitals there is only coupling to the $s$ and $d_{z^2}$ orbitals of 
Pt. In addition, the Hamiltonian matrix elements between these orbitals fulfill 
approximately the condition $H_{1,s} H_{2,d_{z^2}} = H_{1,d_{z^2}} H_{2,s}$, where
1 and 2 stand for the first and second atom of H$_2$. Therefore, the coupling matrix 
between the H$_2$ and the Pt leads is singular, which implies that 
one of the transmission eigenvalues vanishes. This condition of the hopping matrix
elements effectively reduces the rank of the transmission matrix.
In other words, \emph{one of the molecular
states does not couple to the metallic states due to symmetry reasons, reducing
the actual number of channels to one}. This phenomenon of symmetry-inhibited 
modes is well-known in the contexts of light propagation in photonic crystals
\cite{PBG} and sound propagation in periodic structures \cite{ABG}. This symmetry 
induced destructive interference is a unique, but generic feature of the
top position and thus a direct indicator of the relative orientation of the H$_2$ axis to 
the electrodes.

\begin{figure}[t]
\begin{center}
\includegraphics[width=0.85\columnwidth]{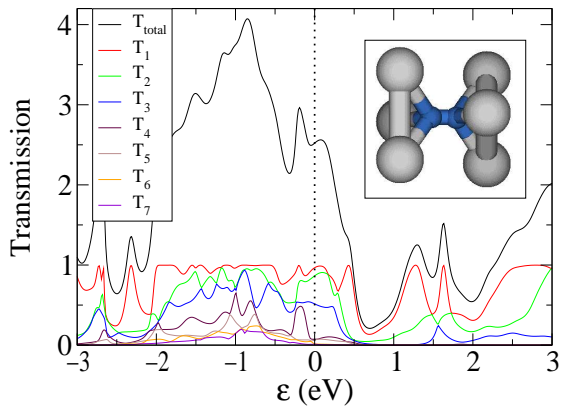}
\caption{\label{Hollow} Transmission as a function of energy for the structure shown
in the inset. The H-H and Pt-H distances are 0.8 {\AA} and 1.86 {\AA} respectively.
The energy of the vibrational mode of the center of mass motion of H$_2$ is 65 meV.  }
\end{center}
\end{figure}

In principle there are other geometries compatible with the vibrational modes
analysis. However, based on the channel analysis performed in the experiment 
many geometries can be ruled out. As an example we consider the situation 
sketched in the inset of Fig.~\ref{Hollow} where each H is bound to three 
Pt atoms (hollow position). Indeed this configuration is suggested in studies 
of the chemisorption of H on Pt surfaces \cite{Kua1998}. The conductance is 
carried by up to 7 individual channels (see  Fig.~\ref{Hollow}). 
Due to the short distance between the Pt leads most of the current is bypassing 
the H$_2$ going directly from Pt to Pt. This analysis allows us to conclude that 
this type of geometries is not realized in the experiment and illustrates the 
importance of the channel analysis.

In conclusion, we have presented a theory for the conductance of a hydrogen
bridge between Pt contacts explaining the experimental observations of Ref.~\cite{Smit2002}.
We have shown how H$_2$ filters out one of the conduction channels of Pt with nearly 
perfect transmission. Our analysis of this ideal test system illustrates that ingredients 
such as charge transfer and the electronic structure of the metallic contacts are 
essential for the proper description of the electrical conduction in single-molecule 
devices. 

We are grateful for many stimulating discussions with R.H.M. Smit, C. Untied
and J.M. van Ruitenbeek. The work is part of the CFN which is 
supported by the DFG. JCC acknowledges funding by the EU TMR Network on 
Dynamics of Nanostructures, and WW by the German National Science Foundation 
(We 1863/10-1), the BMBF and the von Neumann Institute for Computing. 



\begin{thebibliography}{}
\bibitem{Aviram1998}
\emph{Molecular Electronics: Science and Technology}, edited by A. Aviram and
M. Ratner (Annals of the New York Academy of Sciences, New York, 1998).
\bibitem{Joachim2000}
C. Joachim, J.K. Gimzewski, A. Aviram, Nature {\bf 408}, 541 (2000).
\bibitem{Gao2000}
H.J. Gao \emph{et al.}, Phys. Rev. Lett. {\bf 84}, 1780 (2000).
\bibitem{Collier1999}
C.P. Collier \emph{et al.}, Science {\bf 285}, 391 (1999).
\bibitem{Reed2001}
M.A. Reed, \emph{et al.}, Appl. Phys. Lett. {\bf 78}, 3735 (2001).
\bibitem{Metzger1999}
R.M. Metzger, Acc. Chem. Res. {\bf 32}, 950 (1999).
\bibitem{Chen1999}
J. Chen \emph{et al.}, Science {\bf 286}, 1550 (1999).
\bibitem{Cui2001}
X.D. Cui \emph{et al.}, Science {\bf 294}, 571 (2001).
\bibitem{Ventra2000}
M. Di Ventra, S.T. Pantalides and N.D. Lang, Phys. Rev. Lett. {\bf 84},
979 (2000). E.G. Emberly and G. Kirczenow, Phys. Rev. Lett. {\bf 87},
269601 (2001).
\bibitem{Smit2002}
R.H.M. Smit \emph{et al.}, Nature {\bf 419}, 906 (2002).
\bibitem{Reed1997} 
M.A. Reed \emph{et al.}, Science {\bf 278}, 252 (1997).
\bibitem{Kergueris1999}
C. Kergueris \emph{et al.}, Phys. Rev. B {\bf 59}, 12505 (1999).
\bibitem{Reichert2001}
J. Reichert \emph{et al.}, Phys. Rev. Lett. {\bf 88}, 176804 (2002).
\bibitem{Heurich2002}
J. Heurich \emph{et al.}, Phys. Rev. Lett. {\bf 88}, 256803 (2002).
\bibitem{Yaliraki1999}
S.N. Yaliraki \emph{et al.}, J. Chem. Phys. {\bf 111}, 6997 (1999).
\bibitem{Palacios2001}
J.J. Palacios \emph{et al.}, Phys. Rev. B {\bf 64}, 115411 (2001).
\bibitem{DFT}
The DFT calculations have been performed with the code GAUSSIAN98 (Rev. A9),
M.J. Frisch \emph{et al.}, Gaussian Inc. Pittsburgh, PA, 1998.
\bibitem{Mehl1996}
M.J. Mehl and D.A. Papaconstantopoulos, Phys. Rev. B {\bf 54}, 4519 (1996).
\bibitem{Xue2001}
Y. Xue, S. Datta and M.A. Ratner, J. Chem. Phys. {\bf 115}, 4292 (2001).
\bibitem{Brandbyge1999}
M. Brandbyge, N. Kobayashi, and M. Tsukada, Phys. Rev. B {\bf 60}, 17064 (1999).
\bibitem{BP86}
A.D. Becke, Phys. Rev. A {\bf 38}, 3098 (1988). J.P. Perdew, Phys. Rev. B 
{\bf 33}, 8822 (1986).
\bibitem{Ross1990}
R.B. Ross \emph{et al.}, J. Chem Phys. {\bf 93}, 6654 (1990).
\bibitem{note1}
We compute the DOS by projecting into the orthogonal atomic-like orbitals 
obtained via a L\"owdin transformation from the corresponding orbitals of 
the non-orthogonal basis sets used in the DFT.
\bibitem{Cuevas1998} 
J.C. Cuevas, A. Levy Yeyati and A. Mart\'{\i}n-Rodero, Phys. Rev. Lett. 
{\bf 80}, 1066 (1998).
\bibitem{Scheer1998} 
E. Scheer \emph{et al.}, Nature {\bf 394}, 154 (1998).
\bibitem{Nielsen2002}
S.K. Nielsen \emph{et al.}, Phys. Rev. Lett. {\bf 89}, 066804 (2002).
\bibitem{PBG}
W.M. Robertson \emph{et al.}, Phys. Rev. Lett. {\bf 68}, 2023 (1992). 
T.F. Kraus \emph{et al.}, Nature {\bf 383}, 699 (1996).
\bibitem{ABG}
J.V. S\'anchez-P\'erez \emph{et al.}, Phys. Rev. Lett. {\bf 80}, 5325 (1998).
\bibitem{Kua1998}
J. Kua and W.A. Goddard III, J. Phys. Chem. B {\bf 102}, 9492 (1998).
\end{thebibliography}
\end{document}